\documentclass[conference]{IEEEtran}
\IEEEoverridecommandlockouts
\usepackage[utf8]{inputenc}
\usepackage{amsmath,amssymb,amsfonts}
\usepackage{algorithmic}
\usepackage{graphicx}
\usepackage{balance}
\usepackage[draft]{hyperref}
\usepackage{textcomp}
\usepackage{xcolor}
\usepackage{dblfloatfix}
\usepackage{enumitem}

\def\BibTeX{{\rm B\kern-.05em{\sc i\kern-.025em b}\kern-.08em
		T\kern-.1667em\lower.7ex\hbox{E}\kern-.125emX}}
\begin{document}
	
	\title{A Scalable Architecture for Power Consumption Monitoring in Industrial Production Environments
		\thanks{This research is funded by the Federal Ministry of Education and Research (BMBF, Germany) in the Titan project (https://www.industrial-devops.org, contract no.\ 01IS17084B).}
	}
	
	\author{\IEEEauthorblockN{Sören Henning}
		\IEEEauthorblockA{\textit{Software Engineering Group} \\
			\textit{Kiel University}\\
			24098 Kiel, Germany \\
			 \enskip soeren.henning@email.uni-kiel.de \enskip}
		\and
		\IEEEauthorblockN{Wilhelm Hasselbring}
		\IEEEauthorblockA{\textit{Software Engineering Group} \\
			\textit{Kiel University}\\
			24098 Kiel, Germany \\
			\enskip hasselbring@email.uni-kiel.de \enskip}
		\and
		\IEEEauthorblockN{Armin Möbius}
		\IEEEauthorblockA{\textit{IBAK Helmut Hunger GmbH\,\&\,Co.\,KG} \\
			Wehdenweg 122\\
			24148 Kiel, Germany \\
			a.moebius@ibak.de}
	}
	
	\maketitle

	\begin{abstract}
		
		Detailed knowledge about the electrical power consumption in industrial production environments is a prerequisite to reduce and optimize their power consumption. Today's industrial production sites are equipped with a variety of sensors that, inter alia, monitor electrical power consumption in detail. However, these environments often lack an automated data collation and analysis.
		
		We present a system architecture that integrates different sensors and analyzes and visualizes the power consumption of devices, machines, and production plants.
		It is designed with a focus on scalability to support production environments of various sizes and to handle varying loads.
		We argue that a scalable architecture in this context must meet requirements for fault tolerance, extensibility, real-time data processing, and resource efficiency.
		As a solution, we propose a microservice-based architecture augmented by big data and stream processing techniques.
		Applying the fog computing paradigm, parts of it are deployed in an elastic, central cloud while other parts run directly, decentralized in the production environment.
		
		A prototype implementation of this architecture presents solutions how different kinds of sensors can be integrated and their measurements can be continuously aggregated. In order to make analyzed data comprehensible, it features a single-page web application that provides different forms of data visualization.
		We deploy this pilot implementation in the data center of a medium-sized enterprise, where we successfully monitor the power consumption of 16~servers.
		Furthermore, we show the scalability of our architecture with 20,000~simulated sensors.
		
	\end{abstract}
	
	\begin{IEEEkeywords}
		Power Consumption Monitoring, Software Architecture, Microservices, Big Data, Stream Processing
	\end{IEEEkeywords}
	
	\section{Introduction}
	
	Electrical power consumption is a relevant cost component for manufacturing enterprises. Besides economic motives, also legal as well as self-imposed regulations such as ISO~50001 \cite{ISO50001} motivate enterprises to reduce and optimize their power consumption. In particular, load peaks should be reduced as those are significantly more expensive \cite{Albadi2008}.
	
	Due to the immense number of devices, machines, and production plants in such environments, a key	challenge is to identify major consumers.
	Varying and simultaneous workloads on different machines complicate this identification. In order to discover saving potential, it is necessary to monitor all consumers and to visualize and analyze their consumption. The data should be monitored as detailed as possible in order to analyze individual consumers or the consumption at particular points in time intensively. However, also aggregated and pre-configured 
	analyses are necessary to make data comprehensible and to allow for an immediate reaction.
	
	Current trends towards the \textit{Industrial Internet of Things} and \textit{Industry 4.0} bring devices that are increasingly able to monitor their state and resource usage. Equipped with network capabilities, they provide these data to other hardware or software components \cite{Jeschke2017}.
	Combining all sensors into one distributed hardware and software system promises to provide the necessary monitoring infrastructure to optimize power consumption \cite{Shrouf2015}.
	Also older devices that do not offer monitoring mechanisms can be integrated using auxiliary devices such as monitoring power sockets.
	The following features are of particular relevance and should be provided by such a system:
	
	\subsubsection{Data Integration}
	
	Devices and machines in production environments usually come from different manufactures located in different business domains. Furthermore, they are likely to differ in their ages and originate from different generations of technological evolution \cite{VogelHeuser2014}.
	This leads to the situation that also the way they supply data varies widely. Most notably, this is due to the protocols and data formats they use but also to the way they measure. Parameters such as precision, sampling rate, or measurement units may vary from domain to domain.
	In order to compare data of different sensors and to consider the data analysis from a higher level, data first have to be brought into a common format. This also includes converting measurement units or splitting up multiple measurements that are sent together. Moreover, it is likely that not all measurements are of interest and only specific values have to be selected. As the amount of data may be too large to be analyzed, it is often reasonable to first aggregate measurements.
	
	\subsubsection{Data Analysis}
	
	The individual consumption values of devices are often too detailed to draw conclusions about the entire production. Instead, it may often be more reasonable to evaluate data for an entire group of devices. This is even more significant in cases, where devices have more than one power supply, which are monitored individually. It is likely that in such cases, only the aggregated data are of interest.
	
	\subsubsection{Data Visualization}
	
	Visualization of monitored and analyzed data allows a user to draw conclusions about the current state of the overall production. Based on this, a user should be able to make decisions about the further operation.
	
	\subsubsection*{Contribution}
	
	In this paper, we make the following contributions: We define architectural requirements, which such a monitoring infrastructure has to meet in order to be generically applicable for different kinds and sizes of production environments (Section~\ref{sec:requirements}).
	We present an architecture that meets these requirements (Section~\ref{sec:architecture}) and that allows for different ways to deploy it (Section~\ref{sec:deployment}).
	In addition, with our open source pilot implementation\footnote{\url{https://github.com/cau-se/titan-ccp}} (Section~\ref{sec:implementation}), we show how our approach can be deployed in a real production environment (Section~\ref{sec:ibak-deployment}) and we evaluate it in terms of scalability (Section~\ref{sec:scalability}). Finally, we discuss related work in Section~\ref{sec:related-work} and conclude this paper in Section~\ref{sec:conclusions}.

	\section{Architectural Requirements}\label{sec:requirements}
	
	Infrastructures and requirements differ significantly among enterprises and between business sectors. These may change not only from business to business but also within the same application scenario, for example, if after an initial test period additional enterprise departments should be integrated. We aim for an architecture that can be deployed in small-scale production environments as well as in arbitrary large ones. In the following, we describe four key requirements that are of crucial importance for such an architecture.
	
	\subsection{Data Processing in Real-Time while Scaling}
	
	The data transmission, analysis, and visualization in our approach should be performed as quickly as possible in order to allow for an insight into the current infrastructure's status at any time. This is the only way to react to unexpected events or to evaluate the current production process. This requirement needs to be reflected in the architecture design such that, for example, batch processing techniques are not an option for the majority of the analyses.
	
	With a larger production environment, the volume of sensor data increases. This includes both the amount of data per sensor as well as the total number of sensors in the production.
	The requirement for real-time data processing should not be sacrificed if the amount of data increases. In addition, the architecture should also be able to handle varying loads during ongoing operation to avoid downtimes in which the production infrastructure would no longer have been monitored. Besides an increasing load, also a decreasing load should be able to be handled efficiently.
	
	\subsection{Scalability and Resource Efficiency}
	
	If the amount of sensor data grows, more computing power is necessary. To a certain degree this can be achieved by providing more powerful hardware (vertical scaling). However, one quickly reaches a limit where additional computing power can only be accomplished by adding further machines (horizontal scaling). According to Abbott and Fisher's \textit{Scale Cube} \cite{Abbott2015}, horizontal scaling can be obtained in three combinable dimensions: duplicate instances of the software system,	split the managed and processed data, and decompose the software by functionalities.
	Our architecture has to be designed in a way that facilitates an operation on multiple machines and, furthermore, utilizes them efficiently.
	
	The amount of data that is recorded by a sensor is often larger than actually needed for analyses. In order to reduce network traffic and make optimal use of the existing hardware, the sensors (or devices located close to the sensors) should already process as much data as possible.
	However, those edge devices typically operate on limited hardware resources, which are usually not sufficient to execute complex analyses directly on them. Moreover, the given resource capacities are not or only limited extendable and, thus, impede scaling of the software.
	Therefore, an architecture design has to find a balance between optimal resource usage and respecting resource constrains of the edge devices.
	
	\subsection{Scalability and Fault Tolerance}
	
	A horizontally scalable system is inevitably a distributed system whose components communicate via the network. This implies that parts may temporarily become unavailable or fail. Therefore, the software architecture and a corresponding implementation must be designed to tolerate faults and those do not lead to a failure of the overall system.
	Supporting horizontal scaling via duplicating instances also assists in fault-tolerance as failed instances can be replaced by their duplicates.
	
	\subsection{Scalability and Extensibility}
	
	As the number of sensors increases, more data formats and protocols need to be integrated. Moreover, it is likely that large production environments require support for additional metrics. This also applies to analyses and visualizations. An increasing volume of measurement data requires more complex, automatic and, therefore, more domain-specific analyses to make the data understandable. Therefore, the architecture should be designed in an adaptable and extensible way.

	\section{Microservice Architecture}\label{sec:architecture}
	
	Considering the architectural requirements described above, we designed a microservice-based architecture \cite{Newman2015} for the desired monitoring infrastructure.
	Microservice architectures are an approach to modularize software. It divides software into small components, called microservices, that can be used and deployed independently of each other. The separation into microservices is based on business functions. Each service maps to an own business area and provides a complete implementation of it \cite{Hasselbring2017}. This makes it much easier to adapt the component to changing requirements that typically arise from the business area.
	Microservices are isolated from each other. They run in separate processes and do not share any state. Thus, they can independently be started, stopped, or replaced. In particular, microservices can be independently released to production so that a new version of one microservice does not require to update the others. 
	Furthermore, they do not share any implementation or database schema but communicate via transaction-less protocols such as REST.
	This also facilitates an individual choice of programming language, database system and technology stack for each service.
	Loose coupling between microservices enables individual scaling of them and allows the system as a whole to scale more fine-grained \cite{Hasselbring2016}. This avoids wasting computing resources as only those components need to be scaled for which it is necessary.
	Since the individual services only require normal network connections between them, they can be deployed in different contexts. This offers a lot of flexibility in the operation of the software. The main drivers for microservice adoption are, depending on application domain, scalability and maintainability \cite{EMISA2019}. Furthermore, microservice architectures support agile architecture work \cite{SA2018}.
	
	\begin{figure}[bt]
		\centering%
		\includegraphics[width=\linewidth]{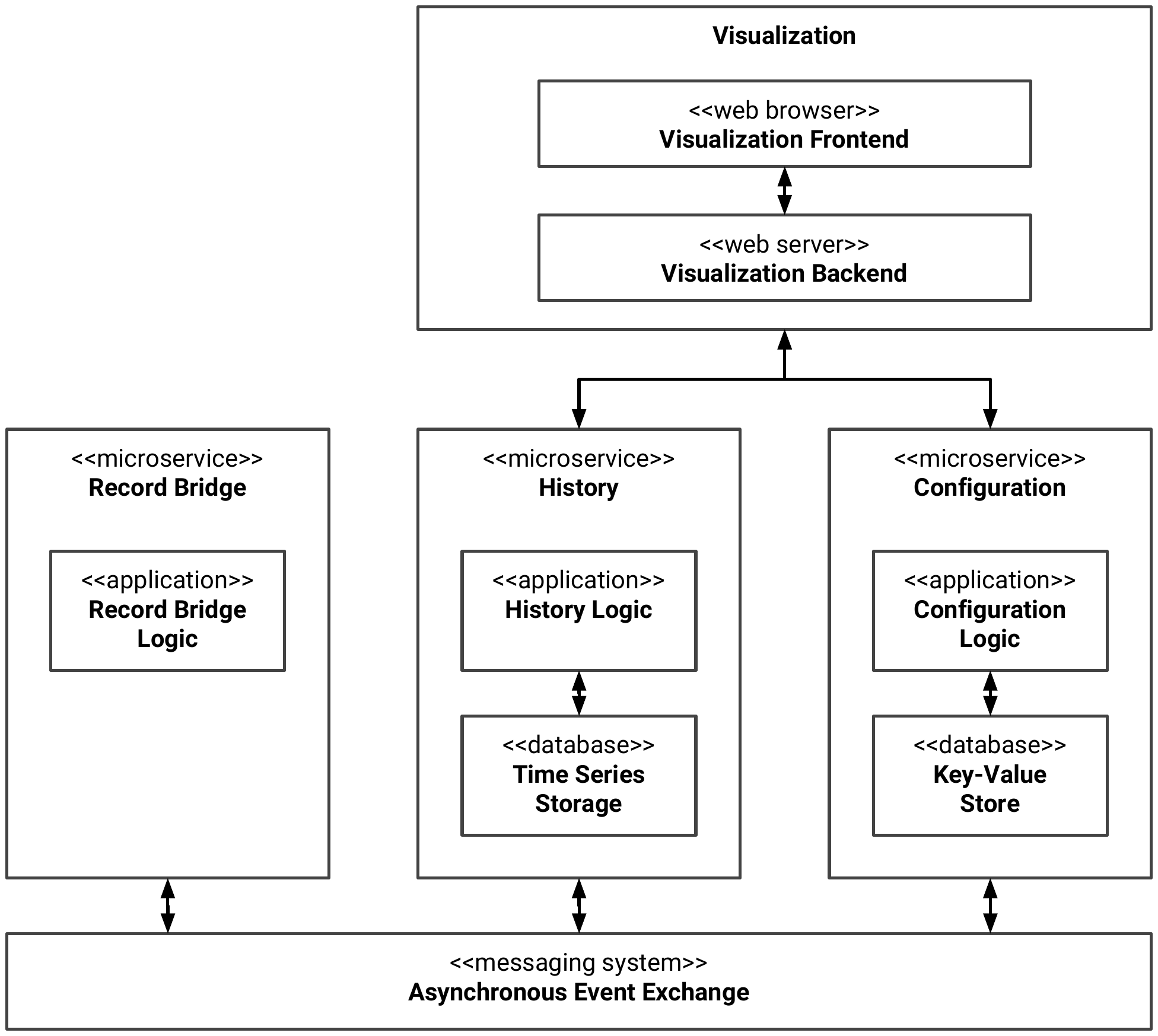}%
		\caption{Microservice-based architecture}
		\label{fig:architecture}%
	\end{figure}%
	
	Fig.~\ref{fig:architecture} shows a graphical representation of our architecture. It contains the three microservices \textit{Record Bridge}, which integrates sensors, \textit{History}, which aggregates and stores sensor data for the long term, and \textit{Configuration}, which manages the system's state. Whereas the Record Bridge solely contains application logic, the services History and Configuration additionally contain a data storage subcomponent. The \textit{Visualization} component is not a typical microservice as it does not represent an own business function but instead serves as an integration of different business functions. It consists of two parts, a server-sided backend and a client-sided frontend.
	
\begin{figure*}[h]
		\centering%
		\includegraphics[width=\textwidth]{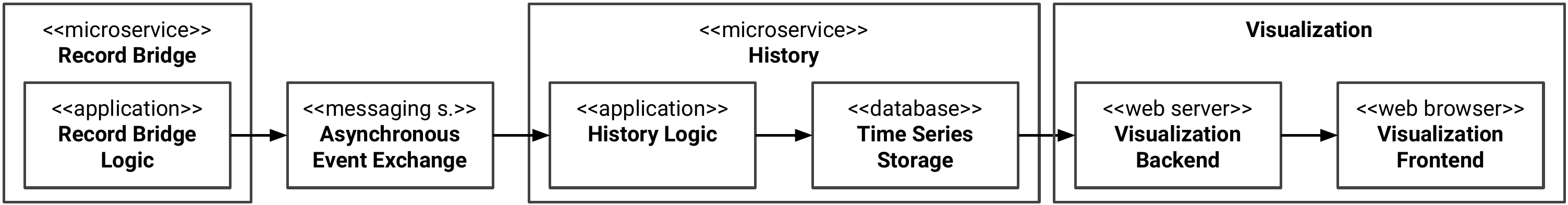}%
		\caption{Data flow between the different components when processing new measurements. The Record Bridge receives monitoring data from physical sensors and queues them into a messaging system. From there, the History service aggregates the data and stores them into a database. The visualization's backend queries that database and forwards the monitoring data to the frontend.}
		\label{fig:architecture-data-flow}%
	\end{figure*}

	The services in our architecture communicate with each other in two different ways: first, synchronously using a request-reply API paradigm such as REST to read or modify the other services' states; second, via a messaging bus or system to publish events that may be asynchronously consumed by other services. Using both communication approaches together is a common pattern when designing microservices \cite{Hasselbring2017}.
	The major task of our approach is stream processing of sensor measurement data.
	Fig.~\ref{fig:architecture-data-flow} shows the flow of measurement data among components starting from their integration via a Record Bridge to their visualization in a web browser.
	
	
	\subsection{Record Bridge}\label{sec:ArchitectureKiekerBridge}
	
	Sensors use different schemata, technologies, and transport mechanisms. Transportation can take place with high-level techniques such as HTTP but also on a low-level, for example, via serial data buses. Data can be encoded in text formats such as JSON or XML but also binary. And besides several standardized data schemata, there are also numerous proprietary ones. This requires to convert sensor measurements to a common format that is used inside our whole approach, before they will be further processed and analyzed.
	
	The Record Bridge fulfills this task. It receives the sensor data, transforms them, and then publishes them for other components by sending them to the messaging system.
	In our architecture design, it functions as a placeholder for arbitrary concrete Bridges, where each Record Bridge integrates a specific sensor type.
	As a sensor type, we consider a set of sensors that use the same schemata, formats, and transport mechanisms.
	
	Record Bridge services are primarily supposed to convert data from one format into another. They do not need to have any or only little knowledge about previous transformations. Therefore, they should be designed as stateless as possible since stateless components enable an arbitrary scaling.
	
	\subsection{History}
	
	The History service manages past sensor data and provides access to them. This includes real sensor measurements as well as aggregated data for groups of sensors.
	Thus, one task this component has to fulfill is the hierarchical aggregation of data. This should be done in realtime, which means: Whenever a sensor supplies a new measurement, all aggregated sensor groups that contain this sensor should obtain an update as well. Like for real sensors, this component creates a new record with the aggregated values and publishes it via the asynchronous event exchange system (bottom Fig.~\ref{fig:architecture}) for other services.
	
	In order to access past measurements, they first need to be permanently stored. Therefore, the History service has access to a database and when it receives new records (aggregated or not) it writes them to that database. For other services, the History service provides access to the database in the form of an API, which has various endpoints that return records or statistics on them.
	
	The application logic is separated from the actual data storage. Thus both parts can be scaled independently. When choosing a database management system (DBMS), it should also be considered how well it can be scaled---both in terms of accessibility and storage. Even if the data retention is segregated into a DBMS, the application logic still cannot be considered entirely stateless. This is due to the fact that multiple instances need to coordinate themselves when consuming data from the messaging system or aggregating them.
	
	\subsection{Configuration}\label{sec:Architecture:Configuration}
	
	The Configuration microservice manages the system-wide settings, such as a hierarchical model that specifies what sensors exist and how they could be aggregated.
	However, the Configuration service does not serve as a central place for all configurations of individual services. Settings that clearly belong to a specific service should be configurable directly in that service.
	An essential requirement of this service is the ability to handle reconfigurations during the execution. In other words, no restart should be needed whenever the configuration changes and other services will receive notifications about those changes. Therefore, the Configuration service provides an API to update or request the current configuration and propagates updates via the messaging system.
	
	Furthermore, this service contains a database to store the current configuration. It is the database's responsibility to store data in a reliable and perhaps redundant manner. Separating the database from the API logic also allows to scale both of them independently.

	\subsection{Visualization}\label{sec:Architecture:Visualization}
	
	
	Besides monitoring and analysis, our approach also includes an interactive, web-based visualization. Our architecture contains a Visualization component following the \textit{Backends for Frontends} pattern \cite{Newman2015}. As the name suggests, it consists of two parts: a \textit{frontend} and a \textit{backend}.
	
	The frontend is a single-page application running in the user's web browser. After compilation, it is a set of static files that are interpreted by the web browser. The actual data are dynamically requested and loaded at runtime from the corresponding microservices.
	
	The backend fulfills two purposes.
	Firstly, it acts as a static file server that delivers the single-page application.
	Secondly, it functions as an API gateway that provides all required interfaces for the frontend.
	When the frontend makes a request, it addresses it to the backend, which then forwards the request to the corresponding microservices.
	In this way, the backend abstracts and hides the internal division into microservices.

	\section{Distributed Deployment}\label{sec:deployment}
	
	
	The proposed software architecture is designed to allow for an individual scaling of its components. In particular, this implies that multiple instances of components can be deployed and that the load is balanced among them. In this way, we expect that our approach is feasible for different sized production environments and, furthermore, we can react flexibly to changing loads and requirements.
	In addition to the software architecture and a corresponding implementation, however, the system must also be deployed in such a way that it can take advantage of the possibilities 
	for scaling.
	
	
	For these reasons, large parts of the architecture are supposed to be deployed in a cloud environment. This does not necessarily have to be a public cloud of an external provider, a private cloud can also offer this. Cloud environments provide the infrastructure and platform demanded by the current load dynamically and \textit{as a service}. This is sensible as hardware in the production is often not powerful enough and provision of additional hardware is time-consuming and costly.
	Therefore, architecture components that perform intensive computations, store data, or operate on the stored data are deployed in the cloud.
	
	However, it may also be reasonable to run particular parts directly in the production environment. Applying the ideas of fog computing \cite{Bonomi2012} and edge computing \cite{Lopez2015},
	we can already reduce the monitoring data where it is recorded.
	That can be achieved by using appropriate filter or aggregate functions. 
	In our architecture design, the Record Bridge can fulfill such tasks but the production environment may also already feature a dedicated edge controller for this.
	Thus, the following four deployment combinations are conceivable (see Fig.~\ref{fig:architecture-deployment}):
	
	\begin{figure*}[h]
		\centering%
		\includegraphics[width=\linewidth]{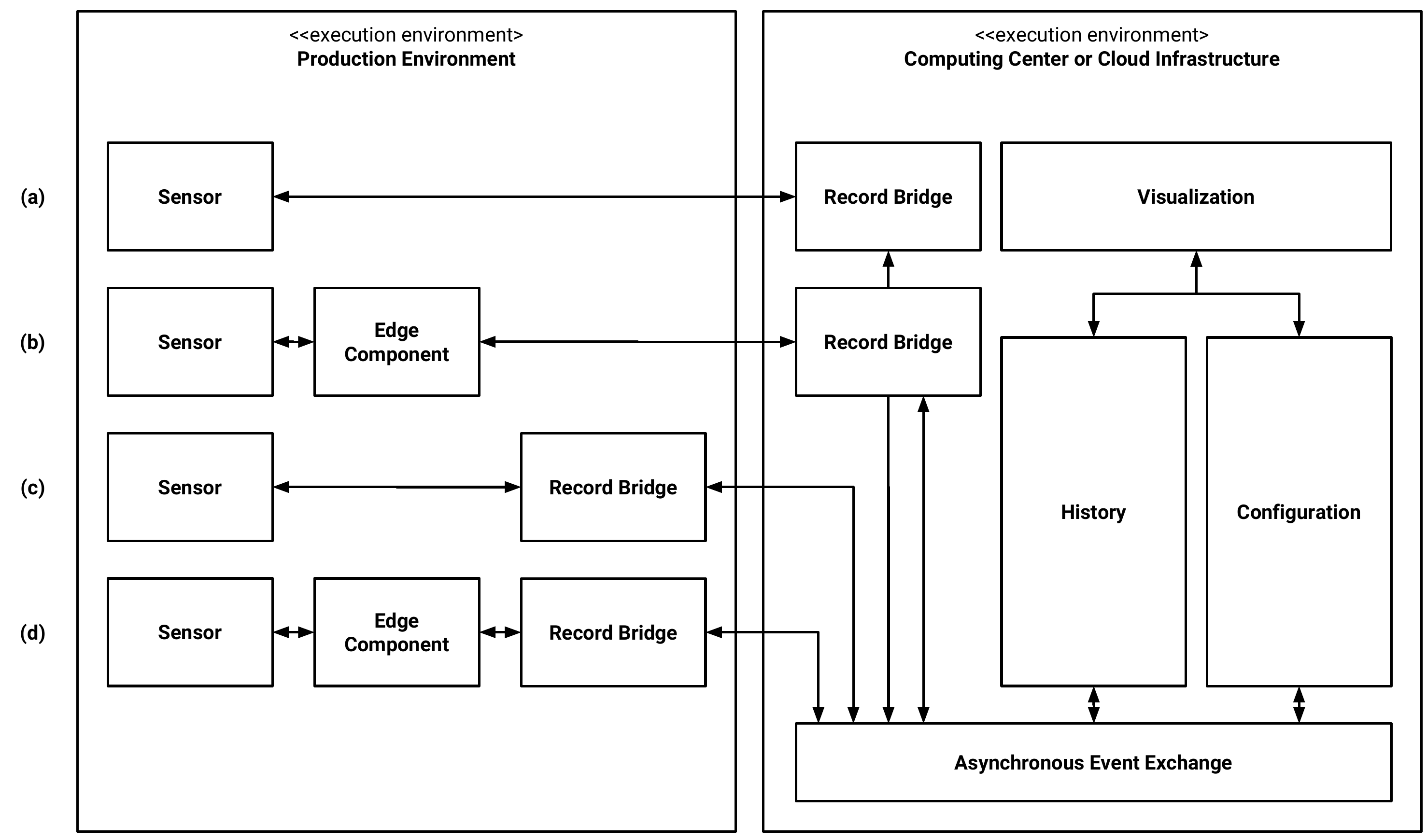}%
		\caption{Deployment architecture showing all possibilities how the Record Bridge and an additional edge component can be deployed.}
		\label{fig:architecture-deployment}%
	\end{figure*}%
	
	\begin{enumerate}[label=(\alph*)]
		\item There exists no separate edge component and the Record Bridge is deployed along with the other microservices in a computing center or cloud infrastructure. This deployment is reasonable if the Record Bridge needs to be scaled dynamically or if there is no appropriate hardware or software infrastructure available in the production environment. Also, this approach is likely to be simpler to realize as its deployment would not differ from the deployment of the other services.
		\item There is a separate edge component and the Record Bridge is deployed in a computing center or cloud infrastructure. This option is reasonable for the same cases as the first one. However, it takes advantage of an existing edge component.
		\item There is no separate edge component but the Record Bridge is deployed in the production environment. In this case, the Record Bridge acts as a kind of edge component that already filters and aggregates monitoring data.
		\item Both an edge component and the Record Bridge are deployed in the production environment. This is perhaps the most future-oriented alternative if hardware gets more powerful and data transmission becomes the limiting factor.
		Depending on the edge controller, it may even be possible to execute both components on the same machine. As data are usually aggregated by the edge component, the Record Bridge solely serves for converting the measurements into a more efficient data format. Only if the aggregation is not configurable enough and an additional filtering of data is necessary, it is reasonable to perform further aggregations by the Record Bridge.
	\end{enumerate}
	
	These approaches can also be arbitrarily combined to adapt to the situation of the existing infrastructure, instead of adapting the production to our approach. Fig.~\ref{fig:architecture-deployment} presents all four approaches within a hypothetical deployment.
	Containerization and orchestration techniques allow to virtualize the execution environment to flexibly assign components to machines.
	

	\section{Pilot Implementation}\label{sec:implementation}
	
	Based on the presented architecture, we developed a pilot implementation of it in the context of our Titan project on Industrial DevOps \cite{Hasselbring2019}. It covers all parts of the architecture including implementations for the individual services as well as the selection of suitable technologies, e.g., databases. In the following, we describe the most important implementation decisions.
	
	\subsection{Communication between Services}
	
	Most services offer REST interfaces, which can be used by other services to request information or to execute operations on them.
	In particular, the visualization in the web browser requests its data via these REST interfaces.
	
	For the asynchronous communication, our implementation uses the messaging system Apache Kafka \cite{Kreps2011}. Kafka can be operated in a distributed cluster of several brokers. Kafka messages consist of a key and a value and are written and read from topics. Topics can be partitioned and the individual partitions are then assigned to one (or, for redundancy, more) brokers. The key of a message is used to assign the message to a partition, which means, messages with the same key are always stored and transferred by the same partition.
	
	Primarily, we use Kafka to transfer sensor measurements. While the message's value is the actual measurement record, we use the identifying name of the corresponding sensor as key. This guarantees that records for the same sensor are always processed by the same Kafka instance, which enhances the scalability for further processing of measurements.
	
	For this prototype, we restrict our implementation to only integrate active power sensor data. Active power records, which we exchange between components, are defined in a data format consisting of an identifier of the sensor, a timestamp, and the measured active power in Watts. Furthermore, we allow to exchange aggregated active power records containing aggregation statistics (e.g., the sum) for a set of records.
	The software performance monitoring framework Kieker \cite{Hoorn2012} offers a domain-specific language (DSL) \cite{Fowler2010} to define such records \cite{Jung2016}. An associated generator creates program code and means to serialize and deserialize records for different programming languages and technologies. We apply Kieker's DSL to define the records.

	\subsection{Integration of Physical Sensors}
	
	The Record Bridge microservices integrate physical sensors by translating the data output of the sensors into the common internal data format. Hence, the architecture envisages a separate Record Bridge microservice for each sensor type. However, the tasks that are fulfilled by those services are largely equal. They have to start the application, load configuration parameters, run continuously, and write records into Kafka topics. They only differ in the way how they receive or request data and how they convert those into Kieker records. Therefore, we provide a Record Bridge framework that eliminates repetitive tasks as much as possible.
	
	The Record Bridge framework considers sensor data as continuous data streams and provides methods to filter and transform these data. A data stream and the operations on it are declaratively described in a Java-based internal domain-specific language (DSL) \cite{Fowler2010}. Using this DSL, one solely has to implement the individual steps that are specific for data formats and technologies. Internally, the stream processing declaration is mapped to a Pipe-and-Filter pipeline, which is interpreted and executed by the framework TeeTime \cite{Wulf2017}.
	
	Similar to other stream processing approaches or functional programming techniques, the source of a stream is a function that generates the elements of it. For example, this can be a web server that creates a stream element for each received HTTP message. A stream can be modified with the following higher-order functions: \textit{filter} retains only specific elements, \textit{map} maps each element of the stream to a new one, and \textit{flatMap} maps each element to multiple new ones. Each of these functions returns a new stream, so that the functions can be concatenated as desired.

	\subsection{Continuous Hierarchical Aggregation}
	
	The History service uses the column-oriented database Apache Cassandra \cite{Lakshman2010} to store records persistently. A web server provides the required REST interface to retrieve the stored data.
	Besides storing and reading, we also require to aggregate measurements of different sensors. One possibility would be to do this when reading records from the database. However, this would be highly computational intensive for frequent queries, in particular, if the records are stored distributed on several nodes. Therefore, we decided to aggregate the data continuously and store the aggregated consumption value along with the real, measured ones. In the following, we describe how the aggregation is computed and how we implemented it in a scalable manner.
	
	\subsubsection{Calculation Methodology}
	
	For an aggregated sensor $\hat{s}$ that should aggregate the sensor group $S = \{s_1,\dots,s_n\}$, its value $v_{\hat{s}}(t)$ at time $t$ is given by the sum of its child sensors' values at that time:
	\begin{align*}
	v_{\hat{s}}(t) = \sum_{s \in S}^{} v_s(t)
	\end{align*}
	
	However, since measured data are only present for discrete points in time, $v_s(t)$ for $s \in S$ is not known for many points in times. Furthermore,  $v_s(t')$ with $t' > t$ is not known since the value should be computed in real-time and thus $t'$ would be in the future. Therefore, it is not possible to perform a simple linear interpolation between the precedent and  successive value. This means in effect, to compute $v_s(t)$ we can only rely on previous values.
	
	For our approach, we equate $v_s(t)$ to the latest measured value. For the interpretation of those data, this means that the time series of the single sensors are shifted towards the future, whereby the shifting interval is at most the temporal distance between measurements.
	If the data sources are measured frequently enough and the values do not fluctuate too much, this procedure should not influence the result notably.
	
	\subsubsection{Realization with Kafka Streams}
	
	\begin{figure*}[t]
		\centering%
		\includegraphics[width=\textwidth]{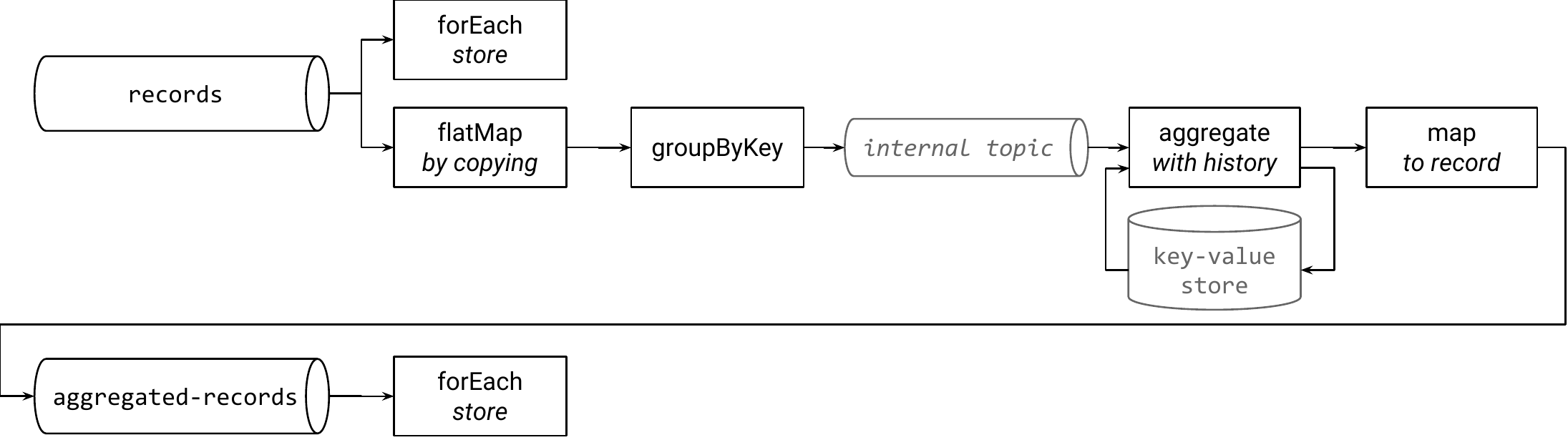}%
		\caption{Graphical visualization of our implemented Kafka Streams topology. Horizontal cylinders represent Kafka topics, the vertical one a database. Grayed-out elements are implicitly created by the framework.}
		\label{fig:kafka-streams-config}%
	\end{figure*}%
	
	In order to implement the calculation methodology described above, we designed a stream processing pipeline using Kafka Streams \cite{Sax2018}.
	Kafka Streams is a stream processing framework build on top of Kafka. In Kafka Streams, processing steps are described in a MapReduce-like manner \cite{Dean2004} to facilitate scalability and fault tolerance. In contrast to MapReduce however, Kafka Streams operates on continuous data streams.
	Fig.~\ref{fig:kafka-streams-config} illustrates this pipeline and pictures the individual steps, which we describe in detail below.
	
	The initial data source is the Kafka topic \textit{records} (top left of Fig.~\ref{fig:kafka-streams-config}). As described above, it contains key-value pairs with a normal active power record as value and its corresponding sensor identifier as key. This topic serves as an interface to the outside of this microservice since it gets its records form other services, namely the Record Bridge services.
	
	Our Kafka Streams configuration consumes the elements of this topic and then forwards them to a \textit{flatMap} processing step. In this step, every record is copied for each aggregated sensor that should consider values of this record's sensor. This means, if a new record is processed, the tree of sensor groups is traversed bottom-up and all parents of the corresponding sensor (parent, grandparent, etc.) are collected in a list. For each entry of this list, the flatMap step emits a new key-value pair with the according parent as key and the active power record as value. 
	
	Those key-value pairs are then forwarded to a \textit{groupByKey} step, which groups records belonging together by serializing them to an internal Kafka topic. Thus, it ensures that all records with the same key are published to the same topic partition and, hence, are processed by the same processing instance in a following step.
	
	The subsequent \textit{aggregate} step maintains an internal \textit{aggregation history} for each aggregated sensor that is processed in the course of time. An aggregation history is a map belonging to an aggregated sensor that holds the last monitoring value for each of its child sensors. It only stores the value for its real child sensors, not for the aggregated ones. Whenever a record arrives with a key for which no aggregation history exists so far, a new one is created. For all successive records the aggregation history is updated by either replacing the last value to this sensor or by adding it if no value for this sensor exists so far. Finally, it is, firstly, stored to an internal key-value store to be used in the next aggregation step and, secondly, forwarded to the next processing step.
	
	Afterwards, the aggregation history is transformed to an aggregated active power record in a \textit{map} step. This is done by calculating different statistics, such as average or sum, of the set of single monitoring values. These aggregated records are then written to the Kafka topic \textit{aggregated-records}. This topic is again designed as an interface such that other services can consume those data, for instance, to perform data analyses on them.
	
	Besides these steps for the hierarchical aggregation, the pipeline also contains two \textit{forEach} steps that asynchronously store the records from both topics \textit{records} and \textit{aggregated-records} to the Cassandra database.
	
	Whereas we declare the single steps of this data processing pipeline, the connection between the steps as well as the serialization to internal topics or databases is handled by Kafka Streams. If multiple instances of this application are started, Kafka Streams manages to balance the data processing subtasks appropriately. A fundamental principle of Kafka Streams is that partitions are always processed by the same instance since in this way no synchronization between reading instances is necessary. Thus, using this approach, we can create as many instances as there are partitions for the \textit{records} and the \textit{aggregated-records} topics. As the number of partitions is bounded by the number of different keys and the keys correspond to the connected sensors, we can start as many instances as there are different sensors and aggregated sensors. This sets a very high limit since the number of sensors will probably be much larger than the degree of parallelization with which the data is processed.
	
	\subsection{Web-based Visualization}

	\begin{figure*}[t]
		\centering%
		\includegraphics[width=\textwidth]{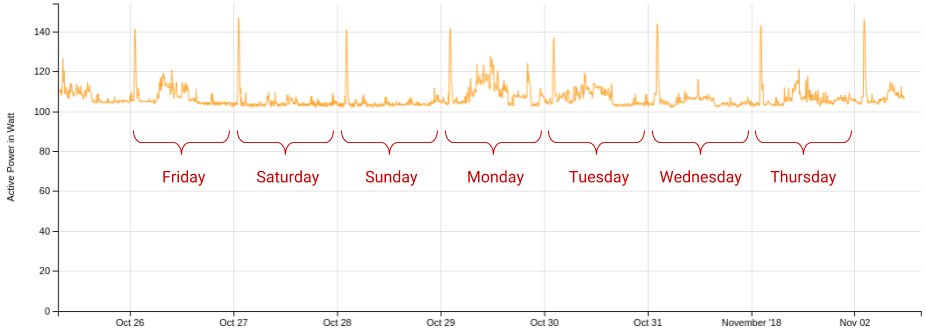}%
		\caption{Power consumption of one of the monitored servers in our pilot deployment.}
		\label{fig:ibak-consumption}
	\end{figure*}%
	
	
	The user interface of the visualization frontend\footnote{http://samoa.se.informatik.uni-kiel.de:8185} is divided into four views (dashboard, sensor details, comparison, and configuration), which can be accessed via the navigation bar on the left side.
	
	The dashboard contains various visualizations of the overall power consumption. In the upper area, it shows three arrows that indicate the trend of consumption in the last hour, 24 hours, or 7 days, respectively. Below them, a large time series chart spans over the entire width. It shows the measured consumption in relation to the point in time it was recorded.
	When new data arrives, the displayed time interval automatically moves forwards. The user can zoom into the chart or move the displayed interval forward and back. Below the time series chart, a histogram shows the frequency distribution of measured values. It serves for recognizing load peaks. Next to the histogram, a pie chart shows the contribution of each sub-consumer. All visualizations update themselves continuously and automatically when new data are available.
	
	The sensor details views is similar to the dashboard view but provides navigation through all consumer and consumer groups so that the consumption of these can be observed in detail.
	The comparison view allows to compare multiple time series interactively. A user can select several time series to be displayed in one chart and, additionally, display multiple charts above each other.
	The configuration view provides a graphical user interface for the Configuration microservice. It allows to add, remove, or rearrange sensors in the hierarchical model via drag and drop.
	
	Research on how to efficiently visualize large data sets was conducted by Johanson et al. \cite{Johanson2016}. 
	In order to provide this visualization, we utilized their library CanvasPlot \cite{Johanson2016Canvas}, 
	which we extended to include real-time functionalities, for our time series charts. Like our other visualizations, CanvasPlot is based on the data-visualization framework D$^3$ \cite{Bostock2011}.
	
	\section{Pilot Deployment}\label{sec:ibak-deployment}
	
	In a pilot deployment, we show that our architecture can be applied to a real industrial environment. For this purpose, we deployed the described prototype in a medium-sized enterprise\footnote{IBAK Helmut Hunger GmbH \& Co. KG}, where we monitored the power consumption of a part of the data center. The deployment includes all parts of our architecture and, thus, covers all aspects of our approach involving data collection, integration, analysis and visualization.
	 
	The monitored part of the data center provides 16~servers that are power supplied by three power distribution units (PDUs). The PDUs have built-in control and monitoring capabilities and can be accessed via the network. Using their embedded web server, we configured them to record the power consumption of each server and push it to a Record Bridge every minute via HTTP. 
	 
	We developed an appropriate Record Bridge that integrates the PDU data using the presented Record Bridge framework. This Record Bridge features an embedded web server that accepts the push messages. A message is encoded in JSON and contains measurements for each PDU outlet, possibly also for several points in time. After receiving the message, the Record Bridge extracts the individual measured values and forwards them as separate records. Furthermore, it only filters the measurements for active power and discards others such as voltage. An aggregation of measurements is not required by this Record Bridge as it is already done by the PDU itself, in our deployment once per minute.
	 
	 
	 
	We run this deployment over a period of three weeks and were able to observe that the measurement data successfully passed through all parts of our approach, from the recording of the PDUs to the visualization in the web browser. Also the operations on the data, such as the continuous aggregation, work as desired.
	 
	Fig.~\ref{fig:ibak-consumption} shows the course of power consumption for one of the 16~servers in a selected time interval of 8~days. The course shows significant nightly peaks at 3~o'clock. Moreover, whereas the power consumption stays at a fairly constant level at weekends, it fluctuates strongly during the day on weekdays and is in average significantly higher than at weekends.
	
	This server is used for desktop virtualization (VDI) for the employees. They mainly work from Monday to Friday, which means that the virtual desktops are used primarily then and remain idle during the weekend. This correlates with the server's power consumption. Every night at 3~o'clock a virus scanner runs on the virtual desktops, which explains the nightly increase in power consumption.


	\section{Experimental Scalability Evaluation}\label{sec:scalability}

	In order to evaluate if the requirement for scalability is met, we examine whether our approach can handle an increasing amount of sensor data with more computing instances. For this scalability evaluation, we simulate a large number of sensors and process their measurements with our prototype implementation. Simultaneously, we measure the number of sensor records processed per second and test this for different numbers of instances. Thus, we determine how many records per second a certain number of service instances can process. As a result, we can determine how many instances are necessary to process the generated load.

	
	We perform this evaluation in a Kubernetes cluster operated in a private cloud infrastructure. It consists of four node servers each featuring 128~GB RAM and $2\times8$ CPU cores that provide 32~threads. The high degree of parallelism allows us to deploy numerous largely independent instances. The nodes and also the experiment are controlled by a dedicated cloud controller.
	
	We developed a simulating Record Bridge that does not integrate external sensors, but generates data itself. For the evaluation, we deployed 20~instances of these Record Bridges. Each of them simulates 1000~sensors that generate one measurement every second so that in total 20,000~records are generated per second.	
	Since the History service is the component that is primarily involved in real-time data processing, we focus on deploying different numbers of History service instances.
	In order to better test parallization characteristics, we limit the computing capacity of each instance to half a CPU core. The Kafka and the Cassandra cluster each consist of three instances. The Kafka topic for the normal active power records contains 20~partitions. For each tested number of History service instances, we determine the average number of processed records per second, repeat this 100~times, and calculate the median as well as the interquartile range of all repetitions. 
	
	\begin{figure}[bt]
		\centering%
		\includegraphics[width=\linewidth]{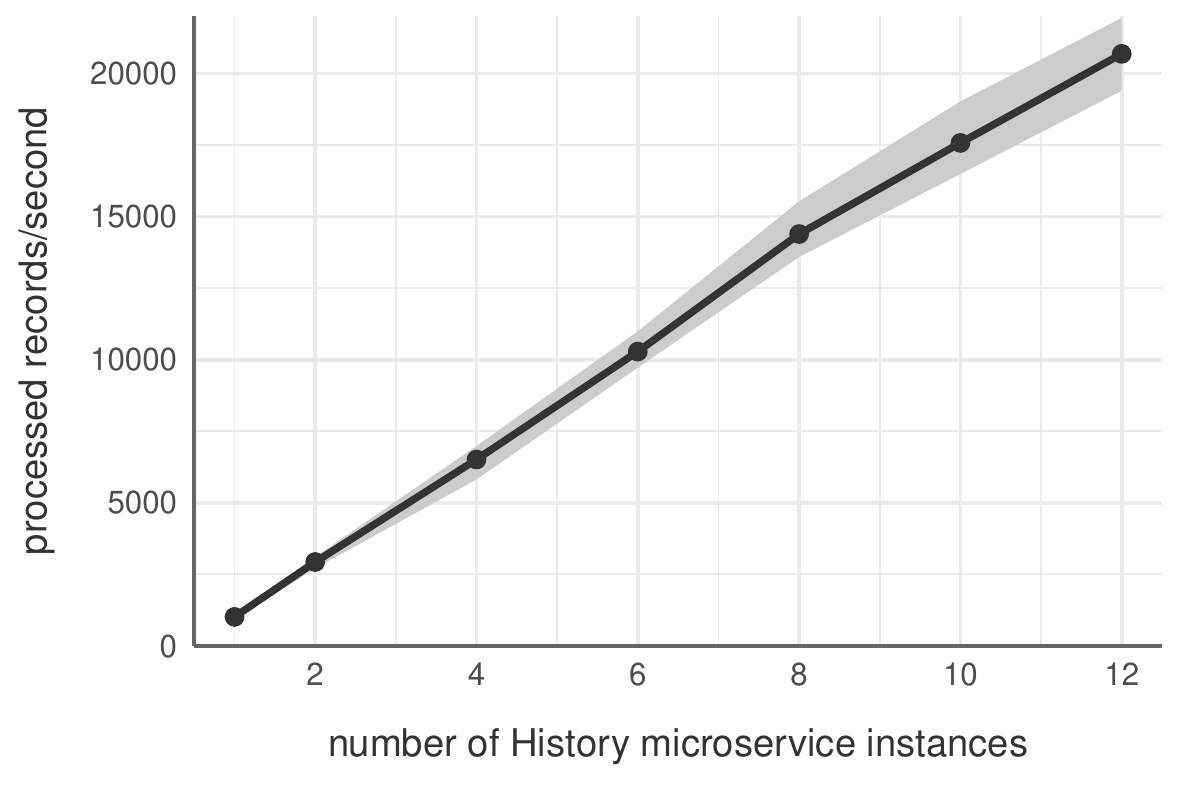}%
		\caption{Amount of processed records per second in relation to the number of processing instances. The median of 100~repetitions is displayed in black. The interquartile range of theses repetitions is highlighted in gray.}
		\label{fig:scalability}%
	\end{figure}%
	
	Fig.~\ref{fig:scalability} shows the amount of processed records per second in relation to the number of processing instances. The amount of processed records scales approximately linear with the number of History service instances. When deploying 12~instances, all measurements that are generated can be processed. Note, as we start the simulation before the processing, values greater than 20,000 are possible. Without the restriction to half a CPU core, significantly higher values would probably be possible since records can be processed faster. During the evaluation we periodically retrieved the CPU and memory utilization of the Kafka and Cassandra instances and verified that the load among instances is balanced evenly. Furthermore, we noticed that API queries (e.g., performed by the visualization) are evenly spread over the History services.
	

	\section{Related Work}\label{sec:related-work}
	
	Power consumption monitoring in production environments is studied by different research disciplines, by industry as well as by academia.

	Shrouf and Miragliotta \cite{Shrouf2015} 
	report on different approaches for energy management enabled by Internet of Things (IoT) technologies. Based on literature, expert interviews, and reports of manufactures, they summarize different IoT architectures for power monitoring and present a general abstraction of them. The resulting architecture primarily focuses on network interconnections and integration of other systems. As in our approach, it respects real-time data processing and the challenge of integrating data of different sensors and data formats. However, data is only processed in a cloud or local server infrastructure and does not follow fog computing paradigms. The architecture represents a general approach and is therefore too abstract to offer a reference implementation.

	A concrete architecture complemented by a prototypical implementation called \textit{Green Cockpit} is presented by Rackow et al. \cite{Rackow2015}. 
	As our architecture, it provides data integration, analysis, and visualization. However, this approach does not consider scalability and real-time data processing to be necessary. Instead, data is integrated by importing text files. Green Cockpit allows for energy planning and other forms of analysis.
	
	
	Sequeira et al. \cite{Sequeira2014} 
	present an energy management system designed for running in the cloud. On the one hand this is, as in our approach, motivated by scalability requirements. On the other hand this approach allows to integrate data of geo-distributed production environments. Whereas our approach relies on fog computing for a decentralized deployment in the production, this approach only supports a centralized deployment in the cloud. 
	The energy management system is based on a Lambda architecture and uses a messaging system to distribute monitoring records. That is, it performs part of the data analyses in real time, while other data are processed in batches. Whereas we designed dedicated isolated microservices for processing, in this approach data is processed in jobs within data processing frameworks.
	
	A comprehensive Industry 4.0 analytics platform is developed at \textit{Bosch} \cite{Groeger2018}. It integrates different kinds of production and business data occurring in an industrial production environment. This platform is also based on a Lambda architecture consisting of a batch, a speed, and a serving layer. As in our approach, scalability is seen as an important requirement and different deployment scenarios are supported. In contrast to our approach, the Bosch approach does not solely rely on open source software but also on commercial ones.

	We did not find any monitoring approaches for production environments designed in a microservices architecture. However, microservice-based approaches exist for other applications of the Internet of Things, such as \textit{smart buildings} \cite{Bao2016} and \textit{smart cities} \cite{Krylovskiy2015}. As we propose in our architecture, these approaches intend to deploy microservices decentralized for flexibility and extendibility. Moreover, they use an asynchronous messaging bus for the exchange of sensor data as in our approach. Both approaches do not focus on scalability and, therefore, do not evaluate this.

	\section{Conclusions and Future Work}\label{sec:conclusions}
	
	Modern industrial production environments offer a number of means to measure resource consumption, such as electrical power, in detail. However, in order to gain knowledge from these data, it is necessary to integrate, analyze and visualize the raw data of the sensors.
	A software and hardware system that provides this in a scalable manner must be designed to a large extent for fault tolerance, extensibility, and efficient resource usage. For useful analyses, data processing should furthermore be carried out in real time.
	
	In this paper, we presented an architecture for such a system that meets these requirements. We apply the microservice architectural pattern that provides solutions to similar challenges in the field of Internet-scale systems. The architecture is intended for a distributed deployment with parts deployed in a cloud environment and parts directly running in the production environment.
	
	For a pilot implementation, we use common technologies for microservices and complement them by techniques and tools for big data processing. We successfully deployed this implementation in the computing center of a medium-sized enterprise and, moreover, were able to show its scalability by simulating 20,000 sensors.
	
	As future work, we plan to supplement our architecture by further microservices. These should primarily provide further and more complex analyses and visualizations, for instance, to automatically detect anomalies in the consumption.
	In order to provide deeper insights into the power consumption of individual production processes, we also work on integrating other consumption metrics as well as production and enterprise data, which can be correlated with electrical power consumption. 
	Furthermore, we plan to conduct extensive evaluations, where we monitor larger production environments with different kinds of devices and machines.
	
	
	\balance
	
	\bibliographystyle{myIEEEtran}
	\bibliography{IEEEabrv,references}
		
	
	
\end{document}